# Tailoring spin waves in 2D transition metal phosphorus trichalcogenides via atomic-layer substitution

Alberto M. Ruiz,[‡a] Dorye L. Esteras,[‡a] Andrey Rybakov[a] and José J. Baldoví[*a]

a Instituto de Ciencia Molecular, Universitat de València, Catedrático José Beltrán 2, 46980 Paterna, Spain.

## Abstract

The family of two-dimensional (2D) van der Waals transition metal phosphorus trichalcogenides has received a renewed interest due to their intrinsic 2D antiferromagnetism, which proves them as unprecedented and highly tunable building blocks for spintronics and magnonics at the single-layer limit. Herein, motivated by the exciting potential of atomic-substitution demonstrated in Janus transition metal dichalcogenides, we investigate the crystal, electronic and magnetic structure of selenized Janus monolayers based on $MnPS_3$ and $NiPS_3$ from first-principles. In addition, we calculate the magnon dispersion and perform real-time real-space atomistic dynamic simulations to explore the propagation of spin waves in $MnPS_3$, $NiPS_3$, $MnPS_{1.5}Se_{1.5}$ and $NiPS_{1.5}Se_{1.5}$. Our calculations predict a drastic enhancement of magnetic anisotropy and the emergence of large Dzyaloshinskii-Moriya interactions, which arises from the induced broken inversion symmetry in the 2D Janus layers. These results pave the way to the development of Janus 2D transition metal phosphorus trichalcogenides and highlight their potential for magnonic applications.

## Introduction

Magnonics is an emerging field of research that deals with the processing, storage and transmission of information by virtue of spin waves (SWs) in magnetic materials. The study of these collective magnetic excitations, whose quanta are called magnons, is currently gaining increasing attention motivated by their higher frequencies, lower energy consumption and efficient tunability, when compared to conventional electronics.[1–4] Since the propagation of SWs does not involve the motion of electrons, magnons are free of Joule heat dissipation, thus offering compelling opportunities toward the fabrication of dense nanoscale devices beyond complementary metal-oxide-semiconductor (CMOS) technology.

In this context, the recent demonstration of long-range magnetic order in van der Waals (vdW) magnetic materials down to the 2D limit[5] provides a particularly interesting platform for magnonic applications.[6] This is due to the combination of low-dimensionality and strong correlation effects as a result of the unique electronic and structural properties of atomically-thin crystals. In addition, 2D materials are characterized by their flexibility and allow a more efficient tuning of their properties with respect to their bulk counterparts. Among the family of 2D magnetic materials, transition metal phosphorus trichalcogenides, of general formula $MPX_3$ (M = transition metal, X = chalcogen, S or Se), add a series of advantages due to their intrinsic 2D long-range antiferromagnetic order,[7–9] air stability[10,11] and the demonstration of long-distance magnon transport over several micrometres in some of their derivatives.[12,13] While typical magnon dynamics in ferromagnets develops at GHz frequencies,[14] the lack of stray fields due to the zero net magnetic moment and their faster dynamics, i.e. faster information processing, with characteristic frequencies in the THz-range, place antiferromagnetic (AFM) materials among the most appealing candidates to propel the frontiers of magnonics.[15,16]

The properties of materials depend intimately on their chemical structures. Over the past few years, tailoring of the optical, electronic and mechanical properties of 2D materials has succeeded through a plethora of chemical and physical approaches such as strain engineering, molecular deposition, external electric fields and intercalation, to name a few.[17–20] This has made them highly versatile for a wide range of applications.[21–24] A well-established approach that drastically modifies the crystal structure of a 2D material is the development of the so-called Janus 2D materials.[25–30] The term Janus refers to structures in which the two faces of the material are asymmetric, leading to out-of-plane broken mirror symmetry. This can result in novel properties such as strong Rashba spin splitting,[31–33] vertical piezoelectricity[34] and second-harmonic generation response.[35] This new class of 2D materials can be fabricated chemically by atomic-layer substitution or asymmetrical functionalization, offering a variety of applications in opto-electronics, electromechanical devices, energy storage, gas sensors, actuators and photo-catalysis.[36–40] Though most of the research in Janus 2D materials come from theoretical simulations, the preparation of Janus graphene by asymmetric functionalization[41,42] and the fabrication of Janus MoSSe monolayer, via both the sulfuration of a single-layer of MoSe$_2$ and selenization of MoS$_2$, has been experimentally achieved.[43,44] Regarding Janus 2D magnetic materials, the first theoretical works have focused on ferromagnetic systems such as transition metal halides and CrSBr.[45,46]

However, despite all this proliferating interest in Janus 2D materials, the influence of their broken mirror symmetry has not been expanded -to the best of our knowledge- to the realm of magnonics. Thus, motivated by the recent experimental advances in the preparation of single-layer Janus materials, herein, we investigate the effect of Janus substitution in transition metal phosphorus trichalcogenides by means of first-principles calculations. We replace the sulfur atoms of one of the surfaces of the MnPS$_3$ and NiPS$_3$ derivatives by selenium, and evaluate their structural, electronic and magnetic properties, as well as their magnon dynamics. Interestingly, we predict that giant antisymmetric magnetic exchange, also known as Dzyaloshinskii-Moriya interactions (DMIs), appears due to symmetry breaking upon Janus modification. This is accompanied by an increase of magnetic anisotropy in the crystal, which gives rise to a dramatically enhanced SW gap in the magnon dispersion that implies a more robust long-range magnetic ordering.

**Computational details**

We performed first-principles calculations using the Quantum ESPRESSO package.[47] We used the generalized gradient approximation (GGA) and the Perdew-Burke-Ernzerhof (PBE) functional to describe the exchange-correlation energy.[48] Standard solid-state ultrasoft pseudopotentials (USPP) from the Quantum ESPRESSO database were selected. In the calculations that include spin-orbit coupling, we used norm-conserving fully-relativistic pseudopotentials from the Pseudo Dojo library.[49] The electronic wave functions were expanded with well-converged kinetic energy cut-offs, where the total energy was converged to 1 mRy/at. All the structures were fully optimized using the Broyden-Fletcher-Goldfarb-Shanno (BFGS) algorithm[50] until the forces on each atom were smaller than $1 \cdot 10^{-3}$ Ry/au and the energy difference between two consecutive relaxation steps was less than $1 \cdot 10^{-4}$ Ry. In order to describe the strong correlations in the $d$ orbitals of the magnetic ions, we adopted a DFT+U approach, where U is the on-site Coulomb repulsion, using the simplified version proposed by Dudarev *et al*.[51] We applied U = 3 eV for Mn and U = 6 eV for Ni based in previous calculations reported in the literature for this family of materials.[52,53] We also tested the influence of the U values on the results in a range 2–4 eV and 4-6 eV, for the $d$ orbitals of Mn and Ni, respectively (see Electronic supplementary information). A fixed 18 Å vacuum space perpendicular to the monolayer plane was used to avoid the interaction with periodic images. The Brillouin zone was sampled by a fine Γ-centered 4 × 3 × 1 k-point Monkhorst–Pack mesh and expanded to 8 × 8 × 1 for magnetic anisotropy energy calculations.[54]

From the Bloch functions obtained by DFT+U, we constructed a tight-binding model based on maximally localized Wannier functions. Our reduced basis set is formed by the $d$ orbitals of the metal centre (Mn, Ni) and the $s$ and $p$ orbitals of P, S and Se. The magnetic interactions were determined using the Green's function method as implemented in TB2J.[55] The one-magnon spectrum was computed using the SpinW code[56] assuming a Hamiltonian up to sixth nearest-neighbours. Group velocities for magnon propagation were calculated by carrying out Landau-Lifshitz-Gilbert (LLG)-based atomistic spin dynamics simulations. We generated SWs by applying an oscillating magnetic field of amplitude 0.13 T and ν = 2 THz during 1 ps to a narrow stripe of the material, which is placed perpendicular to the calculated group velocity direction (parallel to the two main crystallographic axes). An integration step time of 10 fs was selected. We used a 1001×21×1(21×1001×1) supercell along $a(b)$ with periodic boundary conditions along the shortest in-plane side of the sample and free border conditions along the two other directions.

# Results and discussion

Bulk transition metal phosphorus trisulfides, with general formula MPS$_3$, are layered materials stacked by van der Waals interactions that crystallize in a monoclinic lattice with symmetry group C2/m.[57] Each monolayer consists of a honeycomb lattice of magnetic ions, which are connected to six sulfur atoms with trigonal symmetry. These sulfurs are bonded to phosphorus forming S$_3$-P-P-S$_3$ bipyramids that are oriented along the third Cartesian axis. By substitution of one of the atomic-layers of S atoms by Se, we have constructed two Janus 2D monolayers of general formula MPS$_{1.5}$Se$_{1.5}$ (M = Mn, Ni) and fully optimized their lattice vectors and atomic coordinates by means of DFT+U (See Computational details). The crystal structure of the proposed Janus MPS$_3$ is shown in Fig. 1.

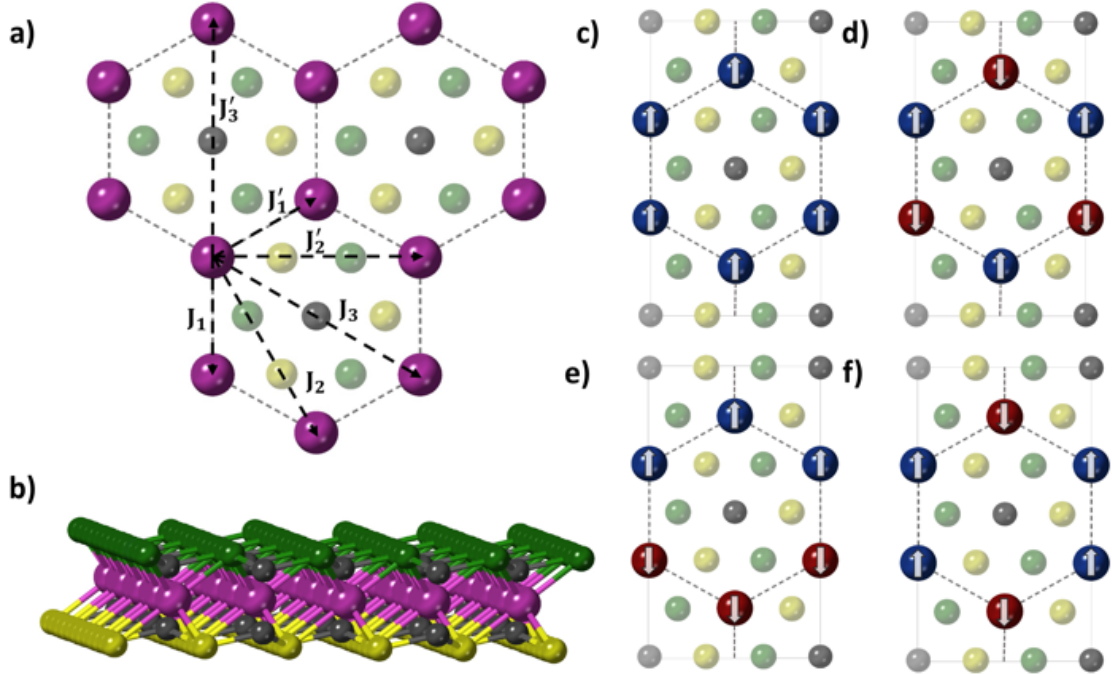

**Figure 1** (a) Top and (b) lateral view of a Janus MPS$_3$ monolayer. $J_1$, $J_1'$, $J_2$, $J_2'$, $J_3$ and $J_3'$ represent the isotropic magnetic exchange parameters in a six-terms model. Top view of different magnetic configurations: (c) FM, (d) AFM-Néel, (f) AFM-Zigzag and (g) AFM-Stripy. Color code: M (purple), P (gray), S (yellow) and Se (green). Blue (red) atoms corresponds to the spin up (down) in the magnetic configurations.

In order to evaluate the effect of the Janus substitution, we determine the electronic structure of the parent compounds, MnPS$_3$ and NiPS$_3$, as well as the selenized monolayers, i.e. MnPS$_{1.5}$Se$_{1.5}$ and NiPS$_{1.5}$Se$_{1.5}$. We tested four possible magnetic configurations in the honeycomb lattice formed by the transition metals, namely, (i) FM, (ii) AFM-Néel, (iii) AFM-Zigzag and (iv) AFM-Stripy (Fig. 1c-f). A rectangular unit cell was used to include all four magnetic orderings. The relative energies scaled to their respective ground state are reported in Table 1. One can observe that our first-principles calculations are in agreement with the experimentally demonstrated AFM-Néel and AFM-Zigzag ground state for MnPS$_3$ and NiPS$_3$, respectively.[58,59] This scenario does not change for the Janus derivatives, as the most stable magnetic configuration remains the same, being also similar in terms of relative energy with respect to other magnetic orderings.

**Table 1** Relative DFT+U energies in (meV/formula unit) for different magnetic orders, namely, ferromagnetic (FM), Néel antiferromagnetic (AFM-Néel), Zigzag antiferromagnetic (AFM-Zigzag) and Stripy antiferromagnetic (AFM-Stripy).

|            | MnPS$_3$ | MnPS$_{1.5}$Se$_{1.5}$ | NiPS$_3$ | NiPS$_{1.5}$Se$_{1.5}$ |
|---|---|---|---|---|
| FM         | 16.79 | 16.96 | 28.56 | 42.57 |
| AFM-Néel   | 0.00  | 0.00  | 7.26  | 12.09 |
| AFM-Zigzag | 7.90  | 6.99  | 0.00  | 0.00  |
| AFM-Stripy | 4.24  | 7.25  | 41.64 | 52.38 |

Regarding structural changes induced by the substitution of S by Se, we observe that in both Mn and Ni cases the lattice parameters get expanded by ~2.6% (Table 2) due to the larger atomic radius of Se atoms with respect to S. This affects the M-M distances ($d_{M-M}$) because of the longer chemical bond between the metal and the selenium atoms (~3.7% and ~4.0% larger than $d_{M-S}$ for M = Mn and Ni, respectively), while the distance M-S remains almost constant. As a consequence, the mirror symmetry is broken in the monolayer, which plays a major role in the enhancement of magnetic anisotropy in the 2D Janus crystals.

Table 2 Structural properties of $MPS_3$ and Janus $MPS_{1.5}Se_{1.5}$ monolayers: Lattice parameters, distances between metallic centres, metallic-chalcogen and chalcogen-chalcogen. All distances are presented in Å.

|  | $MnPS_3$ | $MnPS_{1.5}Se_{1.5}$ | $NiPS_3$ | $NiPS_{1.5}Se_{1.5}$ |
|---|---|---|---|---|
| $a$ | 6.19 | 6.35 | 5.86 | 6.01 |
| $b$ | 10.71 | 10.99 | 10.15 | 10.42 |
| $d_{M-M}$ | 3.57 | 3.67 | 3.38 | 3.46 |
| $d_{M-S}$ | 2.68 | 2.69 | 2.48 | 2.48 |
| $d_{M-Se}$ | - | 2.79 | - | 2.58 |
| $d_{S-S}$ | 3.86 | - | 3.63 | - |
| $d_{S-Se}$ | - | 3.97 | - | 3.80 |

Then, we calculate the electronic band structure with and without spin-orbit coupling (SOC) relativistic effects along the high symmetry points Γ-Z-C-Y-Γ-C (Fig. 2) for the rectangular cell reported in Fig. 1. We obtain bandgaps of 2.2 and 1.85 eV for $MnPS_3$ (direct) and $NiPS_3$ (indirect), respectively, which describe their insulating behaviour. The Janus substitution reduces both of them, providing bandgaps of 1.85 eV and 1.5 eV for $MnPS_{1.5}Se_{1.5}$ and $NiPS_{1.5}Se_{1.5}$, respectively. This is because the larger separation between the metallic ions reduces the energy difference between the bonding and antibonding orbitals around the Fermi level due to smaller overlap of the electronic wave functions. Analogously to $MnPS_3$ and $NiPS_3$, the larger band gap of $MnPS_{1.5}Se_{1.5}$ with respect to $NiPS_{1.5}Se_{1.5}$ is due to four unoccupied bands that appear above the Fermi energy in the latter. These bands correspond mainly to the empty $e_g$ orbitals of $Ni^{2+}$ ($d^8$) due to the crystal field splitting created by the coordination environment, having both spin up and down components of the $t_{2g}$ orbitals fully occupied. However, in the case of $MnPS_3$ and $MnPS_{1.5}Se_{1.5}$, the contribution of the transition metal to the lowest unoccupied molecular orbitals comes mainly from the empty $t_{2g}$ orbitals of $Mn^{2+}$ ($d^5$) ions, which are higher in energy. To provide further chemical insights about the band structure, the role of atomic orbitals is analysed by performing an orbital resolved density of states simulation for each material (Fig. S1-S4). Due to their chemical similarity, it is shown that for all of them the $d$ orbitals of the transition metal play an important role around the Fermi energy as it is characteristic of Mott insulators. In this region, it is also noticeable the high contribution of the $p$ orbitals of S, which also applies for Se in the case of the Janus monolayers. This presence of Se $p$ orbitals in the top of the valence band, i.e. in the highest occupied molecular orbitals, is the responsible of the reduction of the band gap reported in Fig. 2. On the other hand, the $p$ orbitals of P appear on a broader energy range, being less present around the Fermi level. In the four calculated electronic structures the contribution of $s$ orbitals of P is minor and only appears discretely around 3-4 eV.

Now we evaluate the effect of SOC in the electronic band structure, showing that it is almost negligible for $MnPS_3$ and $NiPS_3$ (see inset in Figs. 1a and 1c). However, in both Janus monolayers, a band splitting can be appreciated around the Fermi level (inset in Figs. 1b and 1d). This can be attributed to a possible Rashba SOC effect, as it has been reported in single-layer Janus MoSSe.[33] This momentum-dependent splitting of spin bands has its origin in the breaking of the mirror symmetry of the pristine monolayers by the presence of the Se atoms. As a result, an internal electric field is induced based on electronegativity differences between Se and S atoms, thus generating an electric dipole. Furthermore, as the Rashba spin splitting originates from SOC, its effect increases when decreasing along a group of the periodic table.

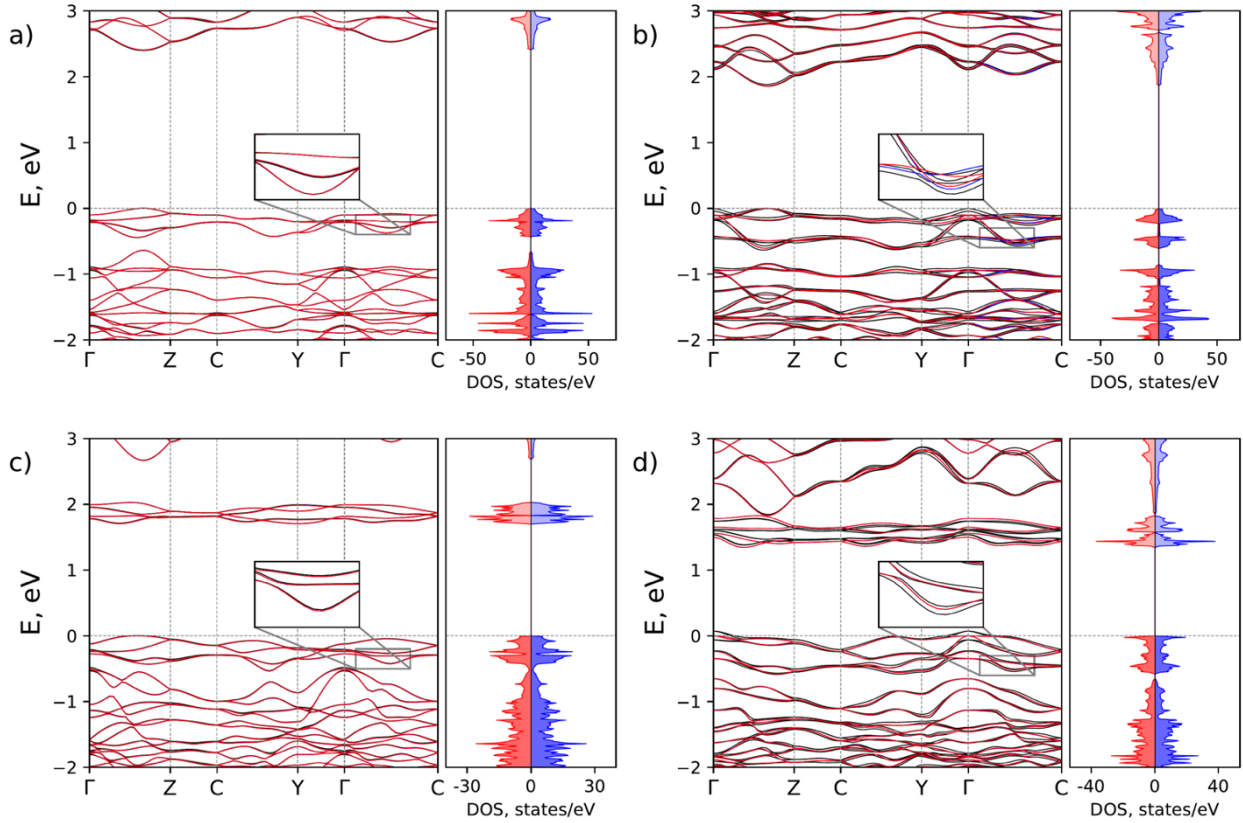

**Figure 2** Electronic band structure and density of states for (a) MnPS$_3$; (b) MnPS$_{1.5}$Se$_{1.5}$; (c) NiPS$_3$; (d) NiPS$_{1.5}$Se$_{1.5}$. Black colour corresponds to DFT+U+SOC calculations; red (blue) colour indicates spin down (up) states in the DFT+U calculations without SOC interaction. Inset: zoom showing the effect of SOC in the electronic band structure.

To provide further insights of the above-mentioned mirror symmetry breaking, we analyse the effect of atomic-layer substitution on the charge density distribution in both materials. We apply a Bader charge transfer analysis, that is an intuitive scheme based on zero flux surfaces to divide atoms, thus computing the electron density flow along the structure in terms of the Bader charges.[60] These results are displayed in Table 3. Focusing on the pristine materials, the analysis reveals that the charge flows from metallic atoms and P to the chalcogenide ligands, as it would be expected by their higher electron affinity. This effect is more pronounced in Mn compounds, where a larger depletion of charge is experienced and can be rationalized according to the lower electronegativity of Mn (1.55) compared with Ni (1.91) in the Pauling scale. On the other hand, the optimized M-S bond distance (Table 2) is ~7.5% smaller in the case of NiPS$_3$ with respect to MnPS$_3$, which implies a more covalent character, being the transition metal less polarized and the chemical bond stronger. Thus, the electron density in the interatomic M-S region grows from Mn to Ni, as it is known,[61] resulting in a smaller depletion of electrons from the transition metal to the ligand, as we observe in the calculated Bader charges. The analysis of the Janus structures reveals that the broken symmetry originated by the S -> Se substitution creates an asymmetric charge distribution. Its origin lies in the fact that the (P$_2$S$_6$)$^{-4}$ bipyramids embedded in the metallic honeycomb structure are now formed by two P atoms coordinated to two different chalcogens. The larger electronegativity of the S atoms with respect to Se causes a more relevant charge depletion in the P atoms that are linked to the sulfur-face of the Janus monolayer. This results in an electron accumulation of 1.72 in the S and 0.48 in the Se ions, which is in agreement with the larger covalent character of the M-Se bond due to the larger size of the anion. The difference between the charge density and the superposition of atomic densities extracted from our DFT calculations is presented in Fig. 3, where blue represents the charge depletion regions and red the accumulation of charge.

**Table 3** Bader charges per atom for MPS$_3$ and MPS$_{1.5}$Se$_{1.5}$. Positive sign indicates depletion of electrons and negative sign electron accumulation.

|        | MnPS$_3$ | MnPS$_{1.5}$Se$_{1.5}$ | NiPS$_3$ | NiPS$_{1.5}$Se$_{1.5}$ |
|--------|----------|------------------------|----------|------------------------|
| Metal  | 1.205    | 1.160                  | 0.784    | 0.696                  |
| P (S)  | 3.894    | 3.915                  | 3.905    | 3.962                  |
| P (Se) | -        | 0.383                  | -        | 0.334                  |
| S      | -1.715   | -1.723                 | -1.571   | -1.589                 |
| Se     | -        | -0.483                 | -        | -0.307                 |

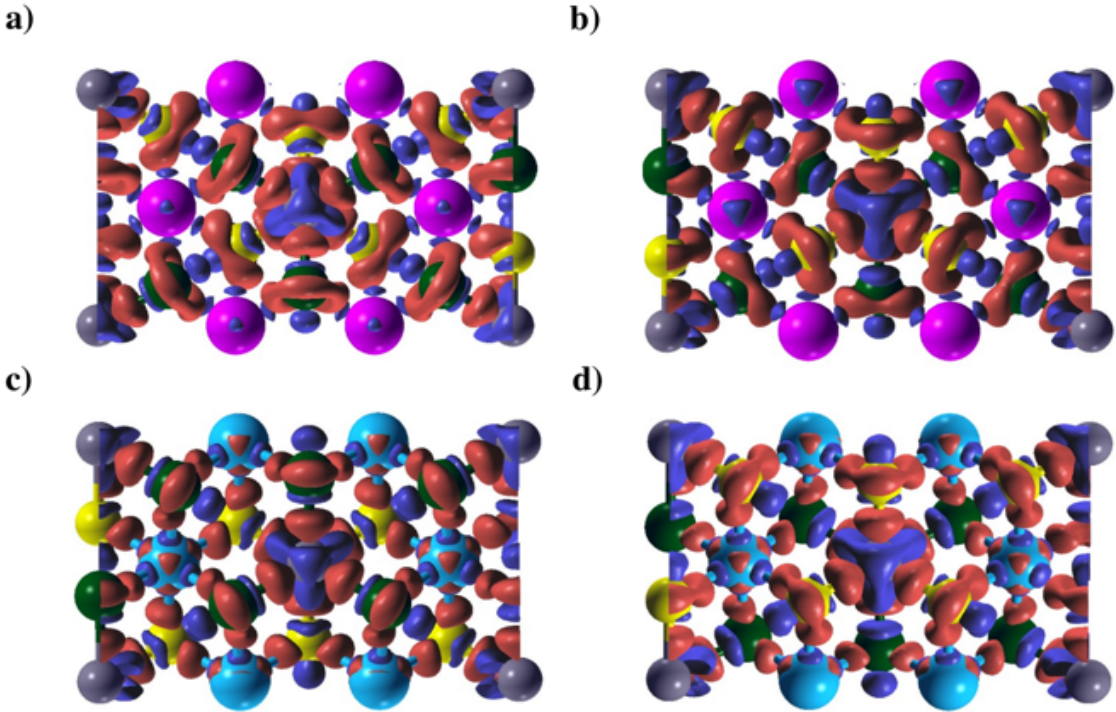

**Figure 3** Valence electronic density differential plot with an isosurface value of 0.007 for the (a) top and (b) bottom view of MnPS$_{1.5}$Se$_{1.5}$; (c) top and (d) bottom view of NiPS$_{1.5}$Se$_{1.5}$. Color code: Mn (pink), Ni (clear blue), P (gray), S (yellow) and Se (green). Blue (red) depicts electron accumulation (depletion) regions.

To properly describe the magnetic properties, we translate the DFT Hamiltonian including SOC expressed in terms of Bloch wave functions into a tight-binding model based on maximally-localized Wannier functions using the Wannier90 code.[62] These real-space localized functions allow us to directly employ the Green's function approach to calculate the magnetic exchange interactions as implemented in the TB2J code.[55] This methodology has been proved to properly describe the properties of 2D magnetic materials and offers an efficient route that avoids the main drawbacks of total energy mapping analysis.[6,63,64] For constructing such Wannier functions, we select the orbitals that play a major role in the electronic properties of the system around the Fermi level. We choose a reduced basis set formed by the *d* orbitals of the metals, and the *s* and *p* orbitals of P, S and Se, in the case of the Janus materials (Figs. S1-S4).

The resulting spin Hamiltonian is shown in Eq. 1:

$$H_{spin} = -\sum_{i} A_i(\vec{S}_i \cdot \vec{e}_i)^2 - \sum_{i \neq j} [\vec{S}_i \mathbf{J}_{ij} \vec{S}_j + \vec{D}_{ij} \cdot (\vec{S}_i \times \vec{S}_j)] \quad (1)$$

where the first term represents the single ion anisotropy (SIA), $A_i$ is the anisotropy parameter and $\vec{e}_i$ is a unit vector along the easy/hard axis direction. The second term indicates the isotropic and symmetric anisotropic (Kitaev) interactions. The last term describes the Dzyaloshinskii-Moriya interaction (DMI), which is the antisymmetric contribution to the global exchange between two spins $S_i$ and $S_j$ and manifests the tendency of spins to be perpendicular to each other. It is important to note that both Kitaev and DMI have their origin in SOC.

The magnetic exchange parameters (second and third term of Eq. 1) are extracted in TB2J as explained in Computational details. The SIA is determined from the magnetic anisotropy (MAE) calculation,[65] which is defined as the energy difference between the magnetic configurations in which spins are aligned along $y$ and $z$ axes. This allows us to obtain a good estimation of the experimentally determined SW gap in NiPS$_3$.[66] Consequently, a negative sign indicates an easy in-plane axis, while a positive sign means an out-of-plane easy magnetization axis.

Due to the symmetry breaking originated by the Janus substitution, SOC highly influences the magnetic properties. Thus, the determination of Kitaev and DMI is key. To compute the exchange parameters, we take into account contributions up to the third nearest neighbour, assuming that there is a non-equal exchange contribution between equivalent nearest neighbours, namely $J_i$ and $J_i'$ (Fig. 1a). This yields a model consisting of six exchange parameters, providing a more accurate description than the conventional three parameters model.[67,68]

Table 4 presents the relative values of DMI and isotropic exchange parameters, describing the competition between both collinear and non-collinear alignment of spins. We can observe that the layer substitution significantly affects magnetic interactions. Moreover, the symmetry breaking shows an important impact at the magnetic anisotropy of the system, and more interestingly, according to our calculations, creates a very large DMI. In the particular case of Ni, we observe a dramatic increase of the |D/J| ratio as a result of the Janus substitution up to 5.3 times. As we have introduced, DMI is by nature antisymmetric. The search for the large DMI appears to be one of the most active topics in condensed-matter physics because it is associated with the formation of exotic topological spin textures such as magnetic helices, skyrmions, antiskyrmions and spirals.[69–72]

**Table 4** Relative strength of isotropic exchange and DMI and MAE (in cm$^{-1}$/atom units)

|  | MnPS$_3$ | MnPS$_{1.5}$Se$_{1.5}$ | NiPS$_3$ | NiPS$_{1.5}$Se$_{1.5}$ |
|---|---|---|---|---|
| $|D_1/J_1|$ | 0.0007 | 0.0263 | 0.0018 | 0.0367 |
| $|D_1'/J_1'|$ | 0.0041 | 0.0260 | 0.0005 | 0.3677 |
| $|D_2/J_2|$ | 0.0445 | 0.1499 | 0.0470 | 0.2999 |
| $|D_2'/J_2'|$ | 0.0445 | 0.1450 | 0.1065 | 0.5643 |
| $|D_3/J_3|$ | 0.0023 | 0.0321 | 0.0003 | 0.0495 |
| $|D_3'/J_3'|$ | 0.0009 | 0.0348 | 0.0002 | 0.0254 |
| MAE | -0.5472 | -2.879 | -0.7895 | 3.3555 |

Furthermore, our MAE calculations indicate that the Janus structure leads to the change of magnetic easy axis from in-plane (NiPS$_3$) to out-of-plane (NiPS$_{1.5}$Se$_{1.5}$). Similar values of MAE for MnPS$_3$ and NiPS$_3$ can be understood from crystal field theory. Both Ni$^{2+}$ and Mn$^{2+}$ possess quenched orbitals and thus their orbital angular momentum (L) is 0. This leads to a null coupling between L and S that results in negligible values of SOC in the magnetic atoms. Hence, the magnetic anisotropy would only be determined by the contribution of the ligands. In Janus structures MAE is dramatically enhanced due to (i) the higher SOC of Se atoms and (ii) the symmetry-breaking-induced anisotropy of the system.

From magnetic exchange Hamiltonian we compute the magnon dispersion using linear spin wave theory as implemented in SpinW package[55] (see Fig 4). As we can observe the Janus substitution (green lines) shift the magnon dispersion to higher frequencies. Furthermore, the larger anisotropy originates an enhancement of the gap around the Γ point. This is especially relevant in the case of the Mn derivative for which it grows from 2.9 to 6.3 meV, resulting in a more robust long-range antiferromagnetic ordering, according to these results. Indeed, we can see that the slope of the magnon dispersion for $MnPS_3$ is decreased due the more pronounced anisotropy gap change with respect to the overall shift of magnon dispersion. This yields to smaller group velocity as reported in Table 5 and it can be understood as the velocity of the propagation of the overall shape of the wave. On contrary, in the case of Ni, the slope of magnon dispersion rises in the Janus derivative at fixed energy since the growth of the gap is comparable with the overall shift of magnon dispersion. Therefore, the group velocity has to enlarge, resulting in larger propagation distance in the material.

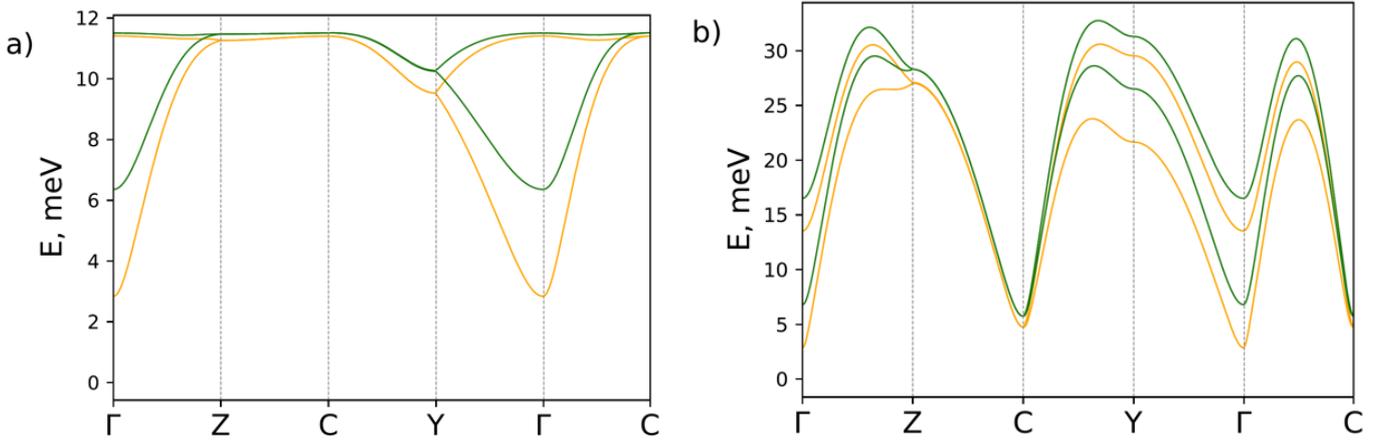

**Figure 4** Magnon dispersion of (a) $MnPS_3$ (orange) and $MnPS_{1.5}Se_{1.5}$ (green); (b) $NiPS_3$ (orange) and $NiPS_{1.5}Se_{1.5}$ (green).

Finally, in order to demonstrate this effect, we perform a computational experiment based on atomistic dynamic simulations. We place two types of thin stripes for each material (along the *a* and *b* crystallographic axis) with an oscillating magnetic field, which is applied along the thin line perpendicular to the longest side of the stripe. This is performed in order to induce a SW as described in Computational details. Note that at the same frequency level (for instance, at 10 meV, i.e. ~2 THz) different magnon modes can be excited as a function of the magnetic centre. In the case of the Mn derivatives high-energy magnons with shorter wave-length can be generated, whereas in the case of the Ni-based ones, long-wavelength magnon modes are induced. The group velocity obtained from our simulation shows that the propagation distance of SWs can be efficiently tuned by Janus substitution, as we predicted, based on the analysis of magnon dispersion (Table 5).

**Table 5** Group velocities along two main crystallographic axis directions for $MnPS_3$, $MnPS_{1.5}Se_{1.5}$, $NiPS_3$ and $NiPS_{1.5}Se_{1.5}$.

|  | $MnPS_3$ | $MnPS_{1.5}Se_{1.5}$ | $NiPS_3$ | $NiPS_{1.5}Se_{1.5}$ |
|---|---|---|---|---|
| $v_a$, $10^3$ m/s | 13.41 | 12.11 | 18.23 | 21.04 |
| $v_b$, $10^3$ m/s | 13.47 | 12.10 | 15.98 | 19.22 |

## Conclusions

In summary, we have investigated by first-principles calculations the structural, electronic and magnetic properties as well as magnon dynamics of the Janus derivatives of single-layer $NiPS_3$ and $MnPS_3$. We prove that the substitution of one S layer by Se generates a spin splitting of the electronic band structures when SOC is included, which derives from breaking of the out-plane symmetry. Indeed, this also induces a very large DMI, a giant increase of magnetic anisotropy energy and an enhancement of the spin wave gap in the 2D Janus $MnPS_{1.5}Se_{1.5}$ and $NiPS_{1.5}Se_{1.5}$. This results in a more robust long-range magnetic ordering, even offering the possibility of switching the easy axis of magnetization from in-plane to out-of-plane direction in the case of the Ni derivatives. Finally, the propagation velocity of the spin waves shows an opposite behavior for Mn and Ni-based materials upon atomic-layer substitution, which is qualitatively independent from the propagation direction. These results pave the road to the development of Janus 2D transition metal phosphorus trichalcogenides and point towards their potential as building blocks for magnonic devices.

## Author Contributions

‡These authors contributed equally to this work. A.M.R. and D.L.E. performed first principles, tight-binding, magnetic exchange and magnon dispersion calculations. A.R. carried out the magnon dynamics simulations and calculated group velocities. A.M.R. and D.L.E. wrote the first draft of the manuscript helped by A.R. J.J.B. conceived the work, proposed the methodology and supervised all the work and the preparation of the manuscript. All authors revised and contributed to the present manuscript.

## Conflicts of interest

There are no conflicts to declare.

*Note*: During the final stage of the review process, we became aware of the work of Jin *et al*.[73] on the preparation of bulk $FePS_{1.5}Se_{1.5}$ from $FePS_3$. This proves the feasibility of the preparation of the proposed Janus materials, since they belong to the same chemical family.

## Acknowledgements

The authors acknowledge the financial support from the European Union (ERC-2021-StG-101042680 2D-SMARTiES and FET-OPEN SINFONIA 964396), the Spanish MICINN (2D-HETEROS PID2020-117152RB-100, co-financed by FEDER, and Excellence Unit "María de Maeztu" CEX2019-000919-M) and the Generalitat Valenciana (grant CDEIGENT/2019/022, CIDEGENT/2018/004 and pre-doctoral fellowship GRISOLIAP/2021/038). The computations were performed on the Tirant III cluster of the Servei d'Informàtica of the University of Valencia.

## Notes and references